\documentclass[preprint,aps,nofootinbib]{revtex4}
\usepackage{graphicx}
\begin{document}

\def\mh{m_h^{}}
\def\gev{\rm GeV}
\def\tev{\rm TeV}
\def\fbi{\rm fb^{-1}}
\def\abi{\rm ab^{-1}}
\def\ee{e^+e^-}
\def\nn{\nu\bar\nu}
\def\ttb{t\bar t}
\def\tth{t\bar t h}
\def\lsim{\mathrel{\raise.3ex\hbox{$<$\kern-.75em\lower1ex\hbox{$\sim$}}}}
\def\gsim{\mathrel{\raise.3ex\hbox{$>$\kern-.75em\lower1ex\hbox{$\sim$}}}}

\hfill$\vcenter{\hbox{\bf MADPH-02-1303}
                 \hbox{\bf UPR--1007T}
                 \hbox{\bf DESY 02--222}
                 \hbox{\bf hep-ph/0301097}}$
\vskip 0.4cm
\title{Effects of genuine dimension-six Higgs operators}
\author{Vernon Barger$^1$\footnote{barger@oriole.physics.wisc.edu}, 
Tao Han$^{1,3}$\footnote{than@pheno.physics.wisc.edu}, 
Paul Langacker$^{1,2}$\footnote{pgl@electroweak.hep.upenn.edu}, 
Bob McElrath$^1$\footnote{mcelrath@pheno.physics.wisc.edu}, 
Peter Zerwas$^3$\footnote{zerwas@mail.desy.de}}
\affiliation{$^1$Department of Physics, University of Wisconsin, 
Madison, WI 53706, USA}
\affiliation{$^2$Department of Physics and Astronomy,
University of Pennsylvania, Philadelphia, PA 19104, USA}
\affiliation{$^3$Deutsches Elektronen-Synchrotron, D-22603 Hamburg, Germany}
\date{\today}

\begin{abstract} 
We systematically discuss the consequences of genuine dimension-six
Higgs operators.  These operators are not subject to stringent
constraints from  electroweak precision data. However, they can modify
the couplings of the Higgs boson to electroweak gauge bosons and, in
particular, the Higgs self-interactions.  We study the sensitivity to
which those couplings can be probed at future $\ee$ linear colliders in
the sub-TeV and in the multi-TeV range.
We find that for $\sqrt s=500$ GeV with a luminosity of 1 ab$^{-1}$
the anomalous $WWH$ and $ZZH$ couplings may be probed to about the $0.01$
level, and the anomalous $HHH$ coupling to about the $0.1$ level.
\end{abstract}


\maketitle

\section{introduction}

Once the Higgs boson is discovered in future collider experiments, the
study of its properties will become the next important goal to
fully establish the nature of electroweak symmetry breaking
\cite{Carena}.  In the Standard Model (SM),  electroweak symmetry
breaking is generated by the Higgs potential
\begin{equation}
    V_{SM}=\mu^2|\Phi|^2 + \lambda|\Phi|^4 \nonumber
      =\lambda{\left(|\Phi|^2-\frac{v^2}{2}\right)}^2 + \, const.  
    \label{vsm}
\end{equation}
for $\mu^2 < 0$. The parameter $v = \sqrt{-\mu^2/\lambda}$, the vacuum
expectation value (vev) of the Higgs field, is determined experimentally
from the well-measured Fermi coupling, $v = 1/\sqrt{\sqrt{2} G_F}$, to
be 246~GeV.  The Higgs boson mass is predicted by the theory to be
$m_H^2=2\lambda v^2$. Once the Higgs boson mass is measured, $\lambda$
is determined and the SM Higgs potential is fully reconstructed.
However, new physics beyond the SM is likely to affect the Higgs sector.
If there are no light degrees of freedom beyond the Higgs boson that
show up in the near future collider experiments, it will be imperative
to scrutinize the interactions of the Higgs boson with the electroweak
gauge bosons as well as the Higgs self-interactions, to search for
evidence of potential extensions of the Standard Model.

In this report we study the consequences of genuine dimension-six
(dim-6) Higgs operators. These operators can be induced by integrating
out heavy massive degrees of freedom in a theory beyond the SM, or they
may reflect the composite nature of the Higgs boson.  Their existence
would provide indirect signatures of new physics that couples
significantly to the electroweak symmetry breaking sector.  Most
naturally, they may be interpreted as the first set of operators in a
series with rising dimensionality of order $1/\Lambda^{2n}$. Therefore
they are expected to be ``small'' with respect to the SM operators so
that the series may be truncated at $dim = 6$ for energies near the
electroweak scale in a first approach to the underlying comprehensive
theory.

Most dim-6 operators involving the SM fermions and gauge bosons
\cite{dim6,wudka} are subject to stringent constraints from the current
precision data \cite{dieter,pdg}. The operators involving the top quark
and Higgs field are less severely bounded, but may be constrained by
theoretical considerations \cite{unit}.
Moreover, they will be studied to a good
precision at hadron colliders \cite{tops} and at future linear colliders
\cite{tt,tth}.  
Those operators involving no other SM fields but the Higgs fields,
\begin{eqnarray}
    {\mathcal{O}}_1 ={1\over 2}
        \partial_\mu(\Phi^\dagger\Phi) \partial^\mu(\Phi^\dagger\Phi)\quad 
    {\rm and}\quad
        {\mathcal{O}}_2 = -{1\over 3} (\Phi^\dagger\Phi)^3,
    \label{dim6text}
\end{eqnarray}
are much less constrained, and they offer potential sensitivity 
to genuinely new structures in the Higgs sector.
The associated Lagrangian is given by the sum
\begin{equation}
    {\mathcal{L}}' = \sum_i^2 {f_i\over \Lambda^2}{\mathcal{O}}_i,
    \label{lp}
\end{equation}
where $\Lambda$ is the energy cutoff at which the new physics threshold
is open and the effective operator approach ceases to be valid. One
naively expects that the scale is near 1 TeV  and that the coefficients
$f_i$ are order of 1 to $4\pi$. Our convention with the relative
negative sign in ${\mathcal{O}}_2$ reflects that this term is induced in
the Higgs potential.

In Sec.~II we study the genuine dim-6 Higgs operators of our current
interest, involving only the Higgs field, and their effects on the
electroweak symmetry breaking. We then present their corrections to the
Higgs boson mass and couplings.  In Sec.~III, we discuss to what extent
the couplings may be probed at $e^+e^-$ linear colliders, such as TESLA
\cite{TESLA} and NLC/JLC \cite{NLCJLC}, with both single Higgs boson and
Higgs boson pair production at energies $\sqrt s=500$ and 800 GeV. In
Sec.~IV, we explore the sensitivity at a multi-TeV collider, as
envisaged with CLIC \cite{CLIC}, for 3 and 5 TeV.  We summarize our
results in Sec.~V. Some details of the analysis for the Higgs sector
with dim-6 operators are presented in an appendix.

\section{Genuine dimension-six operators in the Higgs sector}

At the dimension-six level, operators involving four SM light fermions
are subject to stringent constraints from the current experimental
measurements at low energies \cite{pdg}. Several operators involving SM
gauge bosons and the Higgs field are also constrained from the precision
electroweak data and from the triple gauge boson self-interactions
\cite{wudka,dieter}.  The operators involving the top quark and Higgs
field seem to be less constrained. However, since they are usually
related to anomalous couplings of $Zt\bar t$ and $Wt\bar b$ by gauge
invariance, they will be studied to a high precision when large samples
of top quarks are available at hadron colliders \cite{tops} and future
linear colliders \cite{tt,tth}. Therefore we will not discuss them
further.

There are only two\footnote{
    The operator ${\mathcal{O}}_3 =
    (D_{\mu}\Phi)^\dagger\Phi\Phi^\dagger D^{\mu}\Phi$ affects the
    two-point functions of the $W,Z$ gauge bosons and thus contributes
    to the $\rho$ parameter, leading to $\rho = 1-f_3 v^2/2\Lambda^2$.
    The current constraint from the precision electroweak data
    \cite{pdg} yields a $95\%$ C.L. bound ${|f_3|v^2}/{2\Lambda^2} <
    0.0050$ for $f_3<0$ and $ < 0.0011$ for $f_3>0$.  This implies that
    $ -0.17<f_3\le 0$ or $0\le f_3<0.036$ if $\Lambda=1$ TeV.  Such a
    small coefficient will not result in significant corrections to the
    SM Higgs sector.  We therefore ignore ${\mathcal{O}}_3$ in our
    further analysis and focus on the other two operators.
}
independent operators, ${\mathcal{O}}_1$ and ${\mathcal{O}}_2$
defined in Eq.~(\ref{dim6text}),
at dimension-six that can be constructed solely
from the Higgs field as invariant singlets under all SM gauge
transformations and that are free of constraints from existing
measurements.
In particular, the $\rho$-parameter will not be affected by these two
operators. 
 
Before we study the physical consequences of the operators,
it is interesting to relate this set-up of operators 
to another approach, the Higgs potential
expansion. A general Higgs potential with higher dimensional operators
can be constructed as \cite{pot,Maiani}
\begin{equation}
    V_{eff}= \sum_{n=0} {\lambda_n\over \Lambda^{2n}}
    \left( |\Phi|^2 - {v^2\over 2}\right)^{2+n}.
    \label{peter}
\end{equation}
This expansion systematically includes all higher order terms in the
effective Higgs potential. For instance, at the dim-6 level ($n=1$),
the identification between Eqs.~(\ref{vsm}-\ref{lp}) and (\ref{peter})
is
\begin{equation}
    \mu^2=-(\lambda_0-{3\lambda_1v^2\over{4\Lambda^2}})v^2,\quad
    \lambda=\lambda_0-{3\lambda_1v^2\over{2\Lambda^2}},\quad
    f_2=3\lambda_1.
\end{equation}
The specific form of Eq.~(\ref{peter}) is clearly motivated by the small
field expansion of the potential around the vacuum $v$, which is fixed
at any given order with $V_{eff}=0$.  On the other hand, the general
dim-6 expansion in Eq.~(\ref{lp}) includes derivative operators that
lead to corrections to the kinetic terms and the Higgs-field derivative
couplings.

Higher dimensional operators can arise from integrating out heavy
massive degrees of freedom in theories beyond the SM, or via radiative
corrections.  In weakly coupled theories such as Supersymmetry, the
effects are typically small before reaching the new physics threshold.
The contribution is usually loop-induced and it is suppressed by a
factor of $1/16\pi^2$.  However, if new strong dynamics is involved in
the electroweak sector \cite{hill}, the effects could be quite
appreciable.  In particular, the Higgs self-interactions parameterized
by the dim-6 operators may be significant \cite{bess}.  The effects of
these operators on the triviality and stability of the theory were
studied in \cite{jose}.

\subsection{Electroweak symmetry breaking}

In the presence of dim-6 operators, the modified effective Higgs
potential may be written as
\begin{equation}
    V_{eff}= \mu^2|\Phi|^2+\lambda|\Phi|^4+{f_2\over 3\Lambda^2}|\Phi|^6.
    \label{eq:Veff}
\end{equation}
Electroweak symmetry breaking is obtained via a non-zero vacuum
expectation value of the scalar field.  Minimizing the potential of
Eq.~(\ref{eq:Veff}) with respect to $|\Phi|^2$ leads to the following
expression for the ground state of the field: 
\begin{equation}
    \langle|\Phi|^2\rangle\equiv {v^2\over 2}=
    \frac{\Lambda^2}{f_2}(-\lambda\pm \sqrt{\lambda^2-f_2\mu^2/\Lambda^2}\ ).
    \label{vevtext}
\end{equation}

Interpreting the dim-6 operators as small corrections in the vicinity of
the local minimum of the Higgs potential, the new term in
Eq.~(\ref{eq:Veff}) should be bounded by
\begin{equation}
    { |f_2 \mu^2|\over \lambda^2\Lambda^2 }\ll 1 \quad {\rm or}\quad  
    { |f_2|\over |\lambda|}{v^2_0\over \Lambda^2 } \ll 1,
    \label{eq:llone}
\end{equation}
where $v_0^2\equiv |\mu^2/\lambda|$ is the standard vev of the Higgs
field in the SM limit.  At the same time the signs of the bilinear
coupling $\mu^2 < 0$ and the quartic coupling $\lambda > 0$ must be
fixed such that the Standard Model is reproduced for large $\Lambda^2$.
Moreover, the sign in the solution (\ref{vevtext}) must be chosen
positive for the solution close to the Standard Model. $f_2$ may acquire
either sign. [Alternative assignments and their consequences are
discussed in the appendix.] Expanding the solution in $1/\Lambda^2$, we
find for the ground state of the Higgs field:
\begin{equation}
    {v^2\over 2}\approx 
    {v^2_0\over 2}\left(1-\frac{f_2v^2_0}{4\lambda\Lambda^2} \right)
    \label{caseiitext}
\end{equation}
where we have rewritten the expression in terms of the minimal 
SM vev $v_0^2$. Thus the dim-6 addition to the Higgs potential leads
to a small shift of the Higgs field in the vacuum, decreasing or 
increasing its strength depending on whether the dim-6 part of the 
potential is repulsive or attractive. The value of $v$ is determined
by the Fermi coupling $G_F$ as noted earlier.   

\subsection{Corrections to Higgs boson couplings}
We first note that the dim-6 operator ${\cal O}_1$ affects the kinetic
terms of the Higgs boson propagation, see Eq.~(\ref{O1}) and
(\ref{kin}). Thus we must re-scale the field $\phi$ to define the
canonically normalized Higgs field $H$
\begin{equation}
    \phi = N H,\quad {\rm with}\quad 
    N=\left(1+ {f_1v^2\over \Lambda^2}\right)^{-{1\over 2}}
    \approx 1-{1\over 2}a_1,
    \label{Ntext}
\end{equation}
where 
\begin{equation}
    a_1={f_1v^2\over \Lambda^2}.
\end{equation}
With this field-redefinition, we find for the physical Higgs boson mass 
\begin{eqnarray}
    \label{mhtext}
    m_H^2 &\approx& 2\lambda v^2\left(1 - 
               \frac{f_1v^2}{\Lambda^2} +
               \frac{f_2v^2}{2\lambda\Lambda^2}\right)
    = 2\lambda v^2\left(1 - a_1 + \frac{a_2}{2\lambda}\right), 
\end{eqnarray}
where
\begin{equation}
     a_2={f_2v^2\over \Lambda^2}.
\end{equation} 
Eq.~(\ref{mhtext}) implies that the Higgs mass $m_H$ and the Higgs
self-coupling $\lambda$ are in general independent parameters when new
physics beyond the SM is taken into account.  The mass does not only
depend on the quartic coupling but also on the small but otherwise
arbitrary dim-6 couplings of the generalized potential.  Since the Higgs
boson mass will be directly measured with very high precision, it is
more natural to express all observables in terms of $m_H$ instead of
$\lambda$, as we will do for the rest of the paper.

For the Higgs couplings to the SM particles, the Higgs field is
renormalized by the factor $N$, which results in the
interactions
\begin{eqnarray}
    \label{VVHtext}
    \mathcal{L}_M =
    (M_W^2W^+_\mu W^{-\mu}+{1\over 2}M_Z^2Z_\mu Z^\mu)
    \left((1-{a_1\over 2}){2H\over v}+(1-a_1){H^2\over v^2}\right) \,.
\end{eqnarray}
The Higgs self-interactions can similarly be expressed in terms of
$v,\ m_H$ and the anomalous couplings $a_1,\ a_2$,
\begin{eqnarray}
    \label{Hstext}
    \mathcal{L}_{H^3} =
        -{m_H^2\over 2 v}
        \left( (1-{a_1\over 2}+{2a_2\over 3}{v^2\over m_H^2}) 
        H^3-{2a_1H \partial_\mu H \partial^\mu H\over m_H^2} \right)\\ 
    \mathcal{L}_{H^4} =
        -{m_H^2\over 8v^2} \left( (1-a_1 + {4a_2v^2\over m_H^2}) H^4 -
        {4a_1H^2\partial_\mu H \partial^\mu H\over m_H^2} \right).
\end{eqnarray}
The ${\cal O}_1$ operator, beyond changing the strength of the
self-interactions, introduces derivative couplings of the Higgs field.
They grow with energy and thus will be more significant in a multi-TeV
collider.  The ${\cal O}_2$ operator, however, affects only the strength
of the standard triple and quartic Higgs boson self-interactions.  Some
details are given in the appendix.

\subsection{Experimental probe of the anomalous couplings}

Unlike many studies in the literature which discuss the achievable
accuracies to determine the Higgs couplings in the SM, we stress that
our treatment is a consistent approach to systematically include new
physics effects in the Higgs sector beyond the SM up to the order of
$1/\Lambda^2$.  In addition to the Higgs mass, we have introduced two
new parameters, $a_i=f_i v^2/\Lambda^2\ (i=1,2)$, referred to as {\it Higgs
anomalous couplings}, each of which in turn is related to a model
parameter $f_i$ and the new physics cutoff scale $\Lambda$. If we
naively set the cutoff scale to a value of 1 TeV, we have $a_i \approx
f_i/16$. In the rest of the paper, we explore the range of parameters
\begin{equation}
    -0.5<a_1,\  a_2<0.5,
\end{equation}
which allows $f_i$ to be about $4 \pi$ or less.

The expected Higgs signal cross section depends upon the anomalous
couplings. Assuming the signal cross section can be factorized as a
product of the SM cross section and a factor depending on the anomalous
couplings,
\begin{equation}
    \sigma =F(a_i)\sigma_{SM},
    \label{factor}
\end{equation}
we can relate the accuracy of the cross section measurements to the
change of the anomalous couplings
\begin{equation}
    \Delta \sigma = \frac{\partial \sigma}{\partial a_i} \Delta a_i, \quad 
    {\Delta \sigma \over \sigma} 
        = \frac{\partial F}{\partial a_i} { \Delta a_i\over F}.
\end{equation}
We identify the relative error on the signal cross section measurement
as the statistical uncertainty of the experiment
\begin{equation}
    {\Delta \sigma \over \sigma} = {\sqrt{S+B}\over S},
\end{equation}
where $S=L \sigma=F(a_i)S_{SM}$ is the expected signal events with an
integrated luminosity $L$, $B$ is the (non-Higgs) background.  We thus
determine the error on the anomalous coupling in terms of SM quantities
and the function $F(a_i)$
\begin{equation}
    \label{erra}
    \Delta a_i = \left( \frac{\partial F}{\partial a_i} \right)^{-1}
              \frac{\sqrt{F  S_{SM} /B  + 1}}
                 {S_{SM}/\sqrt{B} } \approx 
              \left(\frac{\partial F}{\partial a_i}\right)^{-1}
             \sqrt{ \frac{F}{ S_{SM}}} ,
\end{equation}
where the last approximation is for zero-background $B$.  The above
estimate is based on simple Gaussian statistics, so the event sample
should be sizable, typically $S\gsim10$ or so.  Also note that a
realistic event determination should include the branching fraction and
experimental efficiency factors for a specific channel, which should be
folded in to $S$ and $B$ here.

Another advantage of our treatment is the factorized formulation as in
Eqs.~(\ref{factor}) and (\ref{erra}). Once the new physics contribution
is formulated as $F$, the sensitivity studies will depend only upon the
SM calculations. In other words, realistic SM signal/background
simulations \cite{TESLA,hwwtesla,Castanier:2001sf,boos-clic} may then
be transformed into a measurement of $a_i$ by Eq.~(\ref{erra}). 

\section{Anomalous Higgs boson couplings at linear colliders}

\subsection{Single Higgs production with anomalous couplings}
\label{Cuts}

\begin{figure}[t]
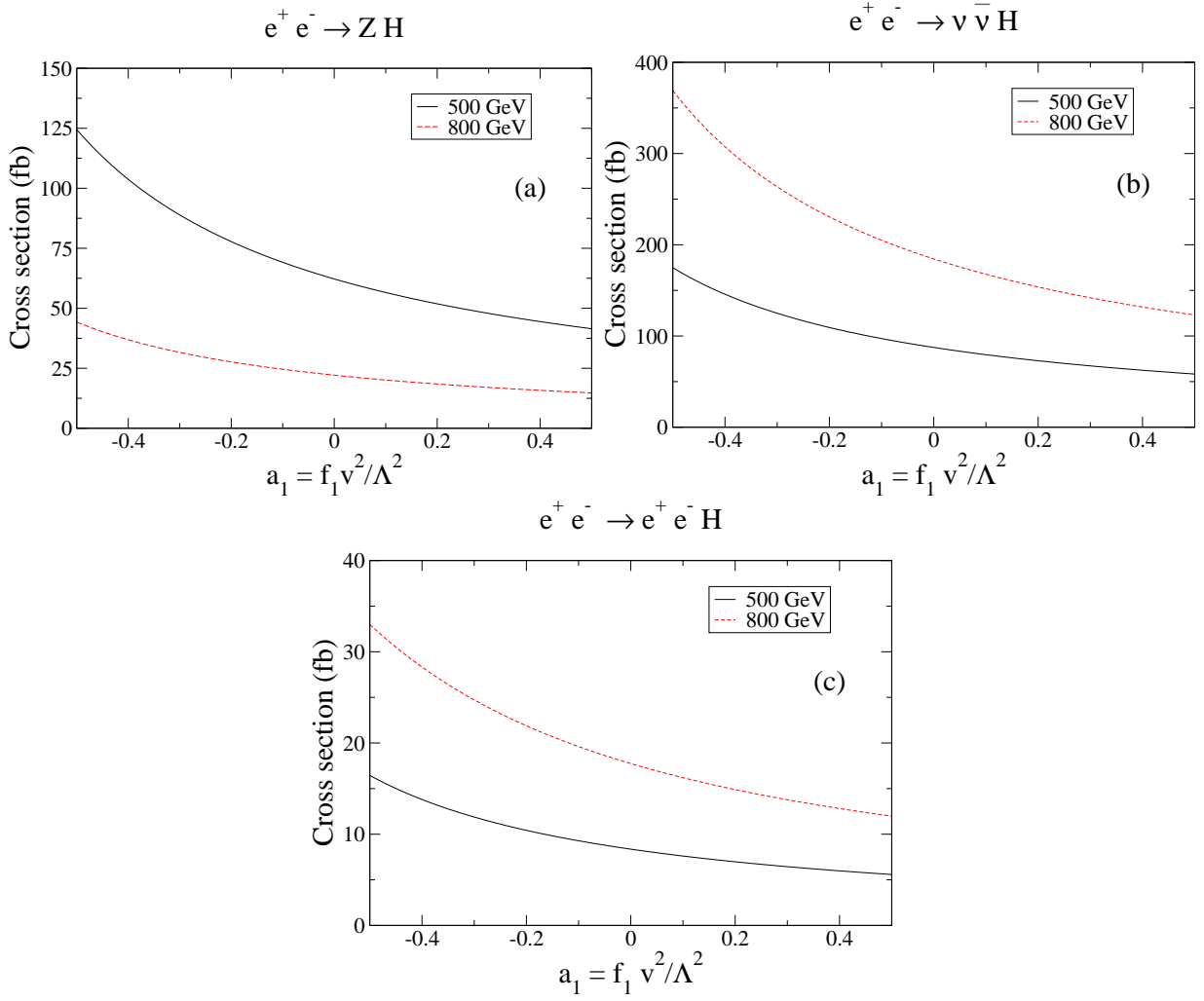

    \includegraphics[scale=0.33]{xsects_zh_a1_LC.eps}
    \includegraphics[scale=0.33]{xsects_nunuh_a1_LC.eps}
    \includegraphics[scale=0.33]{xsects_eeh_a1_LC.eps}
    \caption {Cross sections for single Higgs boson production versus
    $a_1$ at $\sqrt s=500$ and 800 GeV for (a) Higgs-strahlung, (b)
    $WW$-fusion, and (c) $ZZ$-fusion.}
    \label{singleh_lc}
\end{figure}

The processes of single Higgs production include
\begin{eqnarray}
    \label{ZZHeq}
    &&\ee \to ZH,\label{zh}\\ 
    &&\ee \to \nn H  \quad
    {\rm and}\quad 
    \ee H.\label{fuss} 
    \label{WWH}
\end{eqnarray}
The Higgs-strahlung process ($ZH$) dominates for moderate values of the
Higgs mass near the production threshold $\sqrt s \sim M_Z+m_H$, but
falls like $1/s$ at higher energies. The $WW$ and $ZZ$-fusion processes
of Eq.~(\ref{fuss}) take over at higher energies due to the logarithmic
enhancement $\ln^2(s/{M_W}^2) \times \ln (s/{M_H}^2) $; the clean
channel $e^+e^-H$ via $ZZ$-fusion allows the complete reconstruction of
the final state but the cross section is smaller than that from
$WW$-fusion by about an order of magnitude due to the weak electron
neutral-current coupling.  In our treatment, we have separated
the weak boson fusion processes from those involving the resonant $Z$
decay  to $e^+ e^-$ or $\nu \bar{\nu}$. This can be achieved by
appropriate kinematic cuts \cite{hzz}. 

At tree level each diagram in these processes involves exactly one Higgs
vertex, and thus the field redefinition $N$ can be factored out of the
cross section, cf.~\cite{anom}, giving a simple tree level relation for
any single Higgs production process:
\begin{equation}
    \sigma = N^2 \sigma_{SM} = \frac{\sigma_{SM}}{1+a_1}.
    \label{scaling}
\end{equation}

We present the total cross sections\footnote{The evaluation of the 
diagrams has been performed by means of COMPHEP \cite{comphep}.}
of single Higgs boson production in
Fig.~\ref{singleh_lc} for $\sqrt s=500$, 800 GeV and $m_H=120$ GeV versus
$a_1$.  All of the plots display similar behavior.  For negative values
of $a_1$ the cross section increases and for positive values of $a_1$
the cross section decreases while results at $a_1=0$ correspond to the SM
prediction.

\begin{figure}[tb]
    \includegraphics[scale=0.32]{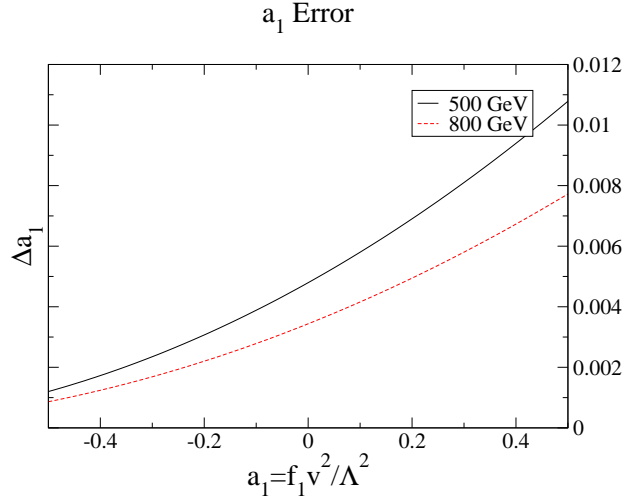}
    \caption {Combined statistical accuracy on $a_1$ with an integrated
    luminosity of 1 ab$^{-1}$ for $\sqrt s=500$ GeV and $800$ GeV, using
    the Higgs-strahlung, the $WW$-fusion and $ZZ$-fusion channels,
    as described in the text.}
    \label{senh_lc}
\end{figure}

We combine four final state channels:  
\begin{itemize}
    \item Higgs-strahlung where  
        $Z\rightarrow \ell^+ \ell^-$ ($\ell=e$ or $\mu$) and Higgs
        decays into anything;
    \item Higgs-strahlung where $Z\rightarrow \nu \bar{\nu}$ and 
        $H \rightarrow b \bar{b}$;
    \item $ZZ$-fusion where Higgs decays into anything;
    \item $WW$-fusion where $H \rightarrow b \bar{b}$.
\end{itemize}
These four channels belong to two well-studied event topologies:
$\ell^+\ell^-$ to reconstruct $m_H$ via the recoil mass while $H$ can
decay into anything; $H\to b\bar b$ plus large missing energy due to the
neutrinos.  For the lepton pair final state, we adopt the acceptance
cuts to identify the Higgs signal 
\begin{equation}
    p_T(\ell)>15\ {\gev}, \quad 
    |\cos\theta_\ell| <0.8,\quad
    M_{recoil}>70\ {\gev}.
\end{equation}
We also impose additional cuts to further remove the backgrounds,
for the Higgs-strahlung
\begin{equation}
    p_T(\ell\ell)>80\ {\gev}, \quad |M_Z-m({\ell\ell})|<10\ {\gev},
\end{equation}
and for the $ZZ$-fusion
\begin{equation}
    m({\ell\ell}) > 100\ {\gev}.
\end{equation}
As for $H\to b\bar b$ plus missing neutrinos, we demand 2-$b$ tag and select
events with
\begin{equation}
    p_T(b)>20\ {\gev}, \quad |\cos\theta_b| <0.8,\quad 
    M_{missing}>70\ {\gev}, \quad |m_H-m({bb})|<10\ {\gev}.
\end{equation}
All of the cuts preserve the signal rate for about $80\%$ efficiencies.
Identification efficiencies of $80\%$ for $b$-tagging and $99\%$ for
each lepton are included \cite{TESLA}.  Although the $\ell^+\ell^-$
final state can reconstruct the Higgs signal very nicely, the $H\to
b\bar b$ from $WW$-fusion and $Z\to \nu\bar \nu$ yields a larger rate
and turns out to be more helpful for the coupling determination.

Given the experimentally observed number of events $S$, along with the
expected SM prediction for the Higgs events $S_{SM}$ and for the
(non-Higgs) background events $B$, the value of $a_1$ and its
statistical error on it can be estimated via Eq.~(\ref{erra}).  We
combine the channels of different topology by quadrature 
to obtain the total error
\begin{equation}
    {1\over \Delta a_2}
      = \left( \sum_j {1\over \Delta a_{2_j}^2} \right)^{1\over 2}.
    \label{sumquad}
\end{equation}
Since the Higgs signal identification by the mass reconstruction in
these channels can be highly efficient at a linear collider,  the
backgrounds are  smaller than the signal rates after the stringent cuts.
We  thus take the zero-background approximation.  Assuming an integrated
luminosity of 1 $\abi$, we plot the combined statistical accuracy on
measuring $a_1$ in Fig.~\ref{senh_lc}.  We find impressive high
sensitivity to those Higgs production processes.  For instance, for
$a_1 = 0$,  $\Delta a_1\approx 0.005$  from Fig.~\ref{senh_lc},
which corresponds to an SM coupling measurement about $0.5\%$.  For
non-zero values of $a_1$, we may reach a typical sensitivity like
$a_1\approx 0.5\pm 0.01$. If we had included the backgrounds in the
analysis, our results on the sensitivity would not change by more than a
factor of $\sqrt 2$. These estimates are close to the results obtained
from detailed experimental simulations, cf. Ref.~\cite{TESLA}.  Further
improved analysis can be converted into an expected measurement of $a_1$
using Eqs.~(\ref{scaling}) and (\ref{erra}).

The energy scale for new physics or composite structure in the Higgs sector
may be inferred from
\begin{equation}
    \Lambda = \sqrt{f_i\over a_i}\ v .
    \label{comp}
\end{equation} 
For a representative coupling $f_1 \approx 1$ and an accuracy $a_1<0.005$,
the resulting energy scale can therefore be bounded to
$\Lambda \geq 3.5$ TeV. 
Thus, these experiments will probe the entire threshold region of
potentially new strong interactions generating the mechanism for
electroweak symmetry breaking. 

\begin{figure}[tb]
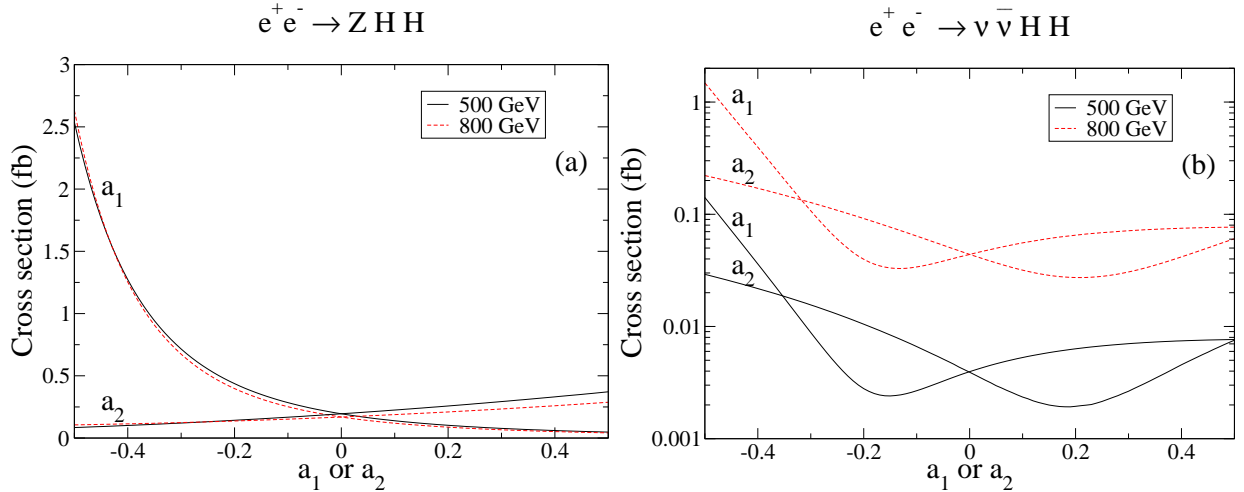

    \includegraphics[scale=0.33]{xsects_zhh_LC.eps}
    \includegraphics[scale=0.33]{xsects_nunuhh_LC.eps}
    \caption {Cross sections for double Higgs production at $\sqrt
    s=500$ and 800 GeV  versus $a_1$ and $a_2$ by (a) Higgs-strahlung,
    (b)  $WW$-fusion.}
    \label{doubleh}
\end{figure}

\subsection{Double Higgs production with anomalous couplings}

The production of two Higgs particles may allow us to probe the
anomalous couplings in $WWHH, ZZHH, HHH$, where the triple Higgs vertex
involves the new operator ${\cal O}_2$ which appears to be rather
secluded from commonly accessible processes. The 
dominant processes that involve
these couplings at linear colliders are \cite{hh,hh2}

\begin{eqnarray}
    \label{ZZhh}
    &&\ee \to ZHH,\\ 
    \label{WWhh}
    &&\ee \to \nn HH.
\end{eqnarray}
The process $\ee \to \ee HH$ via $ZZ$-fusion again is smaller than the
$WW$-fusion  by about an order of magnitude, and thus is not considered
here.  The accuracy of measuring the SM coupling $g_{HHH}$ has recently
been studied at the LC in 
Ref.~\cite{Castanier:2001sf,lc-hhh,jlc-hhh,anom}, and
at CLIC in Ref.~\cite{boos-clic}. Extensions to supersymmetric Higgs
pair production \cite{hh2} and to two-Higgs doublet  models \cite{thdm}
have also been explored in detail.

Inspecting the increasing
order of the electroweak couplings and the increasing number of 
heavy particles in the final states for the 
diagrams of single Higgs-strahlung and double
Higgs-strahlung, it is clear that single Higgs-strahlung will provide
the far better bound or measurement of the anomalous coupling $a_1$.
This parameter is therefore taken as a fixed input, $a_1 = 0$ for
definiteness, for bounding or measuring the new parameter $a_2$ in
double Higgs-strahlung.

The total cross sections versus $a_2$ at $\sqrt s=500$ and 800 GeV are
presented in Fig.~\ref{doubleh} for the double-Higgs-strahlung and
$WW$-fusion.  For completeness the dependence on $a_1$ is also shown 
in the figures.  The contribution from an anomalous triple Higgs coupling that
is derivatively coupled, is essentially proportional to $a_1$.  It
becomes increasingly important only at higher energies.

\begin{figure}[tb]
    \includegraphics[scale=0.32]{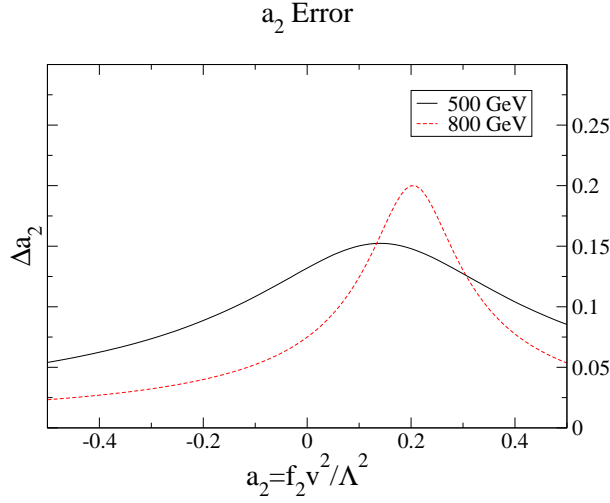}
    \caption {Combined statistical accuracy on $a_2$ with an integrated
    luminosity of  1  ab$^{-1}$ for $\sqrt s=500$ GeV and $800$ GeV,
    using the Higgs-strahlung  and the $WW$-fusion channels, as described
    in the text.  }
    \label{senhh_lc}
\end{figure}

The only contribution from $a_2$ comes from the triple Higgs
self-interaction.  These cross sections are formally quadratic as a
function of $a_2$, resulting from the triple Higgs coupling being linear
in $a_2$. The minimum of the cross section in $a_2$ moves with energy,
and its location in $a_2$ is determined by the size of the crossing of
the triple-Higgs diagram with the other diagrams and itself.  For
instance, in Fig.~\ref{doubleh}(a) of the $ZHH$ plot at 500 GeV the
minimum occurs at $a_2 \simeq -2.4$.  

With our parameterization of the anomalous couplings, the cross section
is quadratic in $a_2$.  We can thus factor out the $a_2$ dependence of
the cross section
\begin{equation}
    \frac{\sigma(a_2)}{\sigma_{SM}} = F(a_2) = A a_2^2 + B a_2 + C,
    \label{abform}
\end{equation}
with the normalization $C=1$. The other coefficients $A,B$ are
fitted to the full calculations and are given in Table~\ref{ab}.  For
the $ZHH$ production, the linear term clearly dominates.  For the
$WW$-fusion process, the linear term is more important for $a_2<0.4$.

The error on the measurement of $a_2$ is determined by Eq.~(\ref{erra}).
The event analysis is similar to that in Sec.~\ref{Cuts}, except that we
require at least 3-$b$ tagging and construct both Higgs mass peaks.  We
combine the errors from the $ZHH$ and $WW$-fusion channels in quadrature
as in Eq.~(\ref{sumquad}).   The statistical accuracy is presented in
Fig.~\ref{senhh_lc}.  We see that for  $a_2 = 0$, we have  $\Delta
a_2\approx 0.1$. For non-zero values of $a_2$, we may reach a typical
sensitivity like $a_2\approx 0.5\pm 0.1$. The worst sensitivity comes
near $a_2\approx 0.2$, where the cross sections reach minimum, see
Fig.~\ref{doubleh}(b).

\begin{table}[tb]
    \begin{tabular}[t]{ l || l ||  c|c}
      Process & $\sqrt s$  & $A$ & $B$  \\
      \hline \hline 
            & 500 GeV & ~0.675~ & ~1.47~   \\
            & 800 GeV & ~0.657~ & ~1.08~   \\
      \raisebox{1.5ex}[0pt]{$Z H H$}
            & 3 TeV   & ~0.374~ & ~0.346~   \\
            & 5 TeV   & ~0.264~ & ~0.243~  \\
      \hline
            & 500 GeV &  ~14.8~  & ~ $-$5.51~  \\
            & 800 GeV &  ~8.84~  & ~ $-$3.66~  \\
      \raisebox{1.5ex}[0pt]{$\nu \bar{\nu} H H$}
            & 3 TeV   &  ~3.21~ & ~ $-$1.69 ~ \\
            & 5 TeV   &  ~2.42~ & ~ $-$1.34~  \\
    \end{tabular}
    \caption{Parameters $A,B$ as in Eq.~(\ref{abform}). 
    \label{ab} }
\end{table}

Since there are a number of studies on the triple Higgs coupling, we can
convert the usual $\delta g_{HHH}/g_{HHH}$ to our $\Delta a_2$ 
at $a_2=0$ and compare with our results.  
From Eq.~(\ref{Hstext}), we have, for $m_H=120$ GeV and
$a_2=0$,
\begin{equation}
    \frac{\delta g^{}_{HHH}}{g^{}_{HHH}} = \frac{2 v^2 
    \delta a_2}{3 m_H^2 + 2 v^2 a_2}\approx 2.8 \Delta a_2.
\end{equation}
Using the results of Ref.~\cite{Castanier:2001sf} on $\delta g_{HHH}^{}/g_{HHH}^{}$
at 500 GeV, we obtain the estimated accuracy on  $\Delta a_2$
\begin{center}
    \begin{tabular}[t]{c|| c | c | c}
        Luminosity   & 500 fb$^{-1}$ & 1 ab$^{-1}$ & 2 ab$^{-1}$ \\
        \hline        \hline
        $\delta g_{HHH}^{}/g_{HHH}^{}$ & 42\%       & 30\%       &  20\%\\
        \hline
        $\Delta a_2$ & 0.15       & 0.11       & 0.073
    \end{tabular}
\end{center}
This is consistent  with our  results for $a_2=0$ 
given in Fig.~\ref{senhh_lc}. Note that 
for ${a_2 \leq 0.073}$, the bound corresponds to an energy scale of 
${\Lambda \geq 910\, {\rm GeV}},$ assuming $f_1=1$ and $m_H=120$ GeV.

\section{Anomalous Higgs boson couplings at CLIC}

At higher energy colliders such as CLIC \cite{clic}  with c.m.~energies 
of $\sqrt s=3$ to 5 TeV, it is quite possible that new physics
thresholds may open and new massive particles be produced.
In the pessimistic scenario that  there are no new particles produced,
higher energy colliders will probe new physics 
phenomena indirectly at a higher 
energy scale $\Lambda\gsim  \sqrt s$. In particular, 
energy-dependent operators will have significant enhancements,
while the $s$-channel SM processes will be suppressed at higher
energies.

\subsection{Higgs production with anomalous couplings}

\begin{figure}[tb]
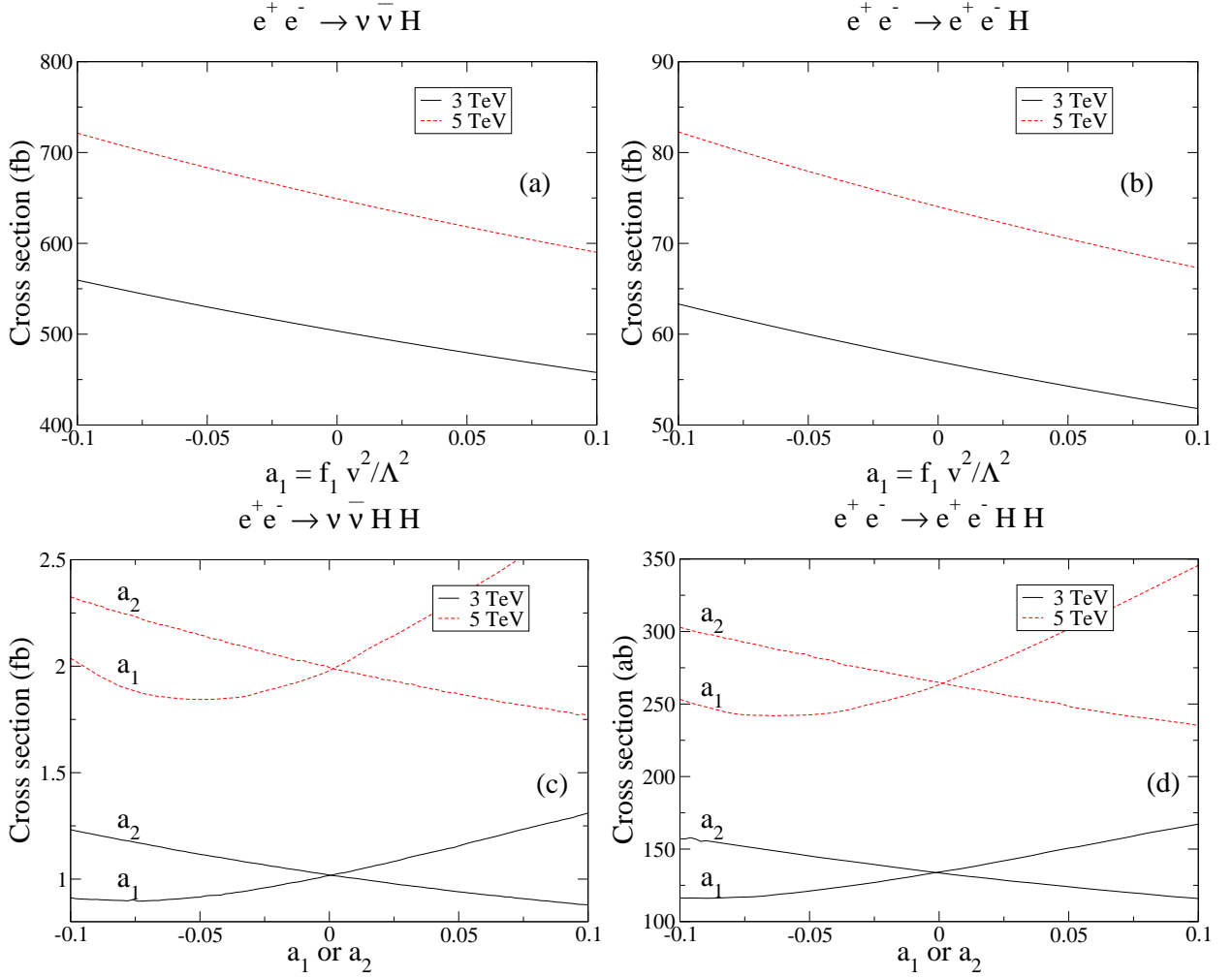

    \includegraphics[scale=0.33]{xsects_nunuh_a1_CLIC.eps}
    \includegraphics[scale=0.33]{xsects_eeh_a1_CLIC.eps}
    \includegraphics[scale=0.33]{xsects_nunuhh_CLIC.eps}
    \includegraphics[scale=0.33]{xsects_eehh_CLIC.eps}
    \caption {Cross sections at $\sqrt s=3$ and 5 TeV 
(a) for single Higgs boson production versus $a_{1}$ via
 $WW$-fusion and (b) via $ZZ$-fusion;
(c) for double Higgs boson production versus $a_{1,2}$ via
 $WW$-fusion and (d) via $ZZ$-fusion. }
    \label{singleh_clic}
\end{figure}

\begin{figure}[tb]
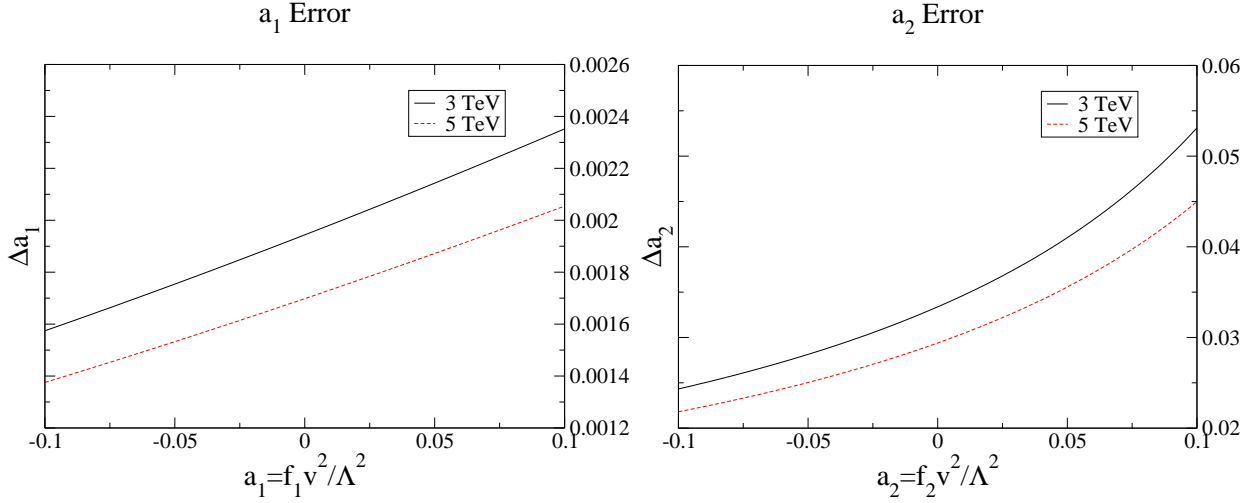

    \includegraphics[scale=0.32]{err_h_a1_CLIC.eps}
    \includegraphics[scale=0.32]{err_hh_a2_CLIC.eps}
    \caption {Combined statistical accuracy on (a) $a_1$ and (b) $a_2$
    with an integrated
    luminosity of  1  ab$^{-1}$ for $\sqrt s=3$ and $5$ TeV,
    using the $WW$- and $ZZ$-fusion channels, as described
    in the text.  }
    \label{senhh_clic}
\end{figure}

We first show the total cross sections for single and double
Higgs production
at the proposed CLIC  energies versus the anomalous couplings
$a_1$ and $a_2$ in Fig.~\ref{singleh_clic}, where we have only
included the dominant channels via $WW$ and $ZZ$ fusion.
It is obvious that due to the much larger production rate, $a_1$
can be probed better via the single Higgs production processes.
For Higgs pair production, 
the processes are again more sensitive to $a_2$ than $a_1$.
The statistical accuracy on $a_1$ can be estimated similarly to
the case of low energies. 
We adopt the same cuts as in Sec.~\ref{Cuts}.
We see from Fig.~\ref{senhh_clic} that the statistical sensitivity of probing 
$a_1$ and $a_2$ at CLIC is improved by roughly a factor of two to 
three over an LC. However, part of the improvement will be lost 
as a result of increased experimental complexity owing to the rapidly
rising beamstrahlung.

\subsection{The triple-Higgs derivative coupling}

\begin{figure}[tb]
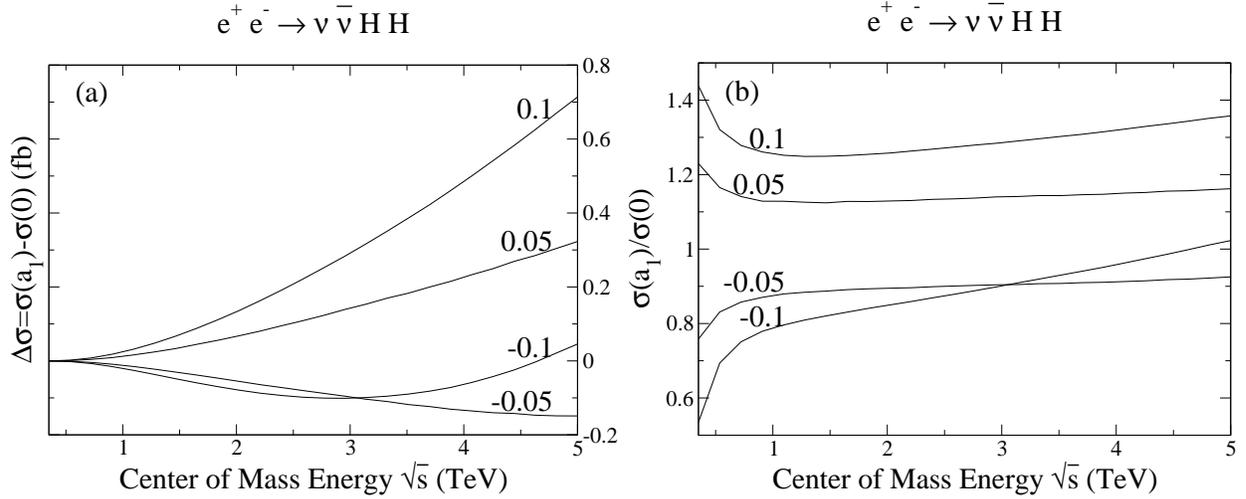

    \includegraphics[scale=0.33]{xsect_nunuhh_roots.eps}
    \includegraphics[scale=0.33]{xsect_ratio_nunuhh_roots.eps}
    \caption{
       (a) Cross section difference and (b) ratio to the Standard Model cross
        section for double-Higgs production versus energy in $WW$-fusion
        for representative values  $a_1=-0.1,\ -0.05,\ 0.05,\ 0.1$.
    }
    \label{Edep}
\end{figure}

The increasing fusion cross sections for single Higgs production 
with rising energy, cf. Fig.~\ref{singleh_clic}, may balance to some
extent the loss of clarity in a final state involving neutrinos.
The new point at rising energy however will be to probe 
the derivative contribution to the anomalous tri-linear coupling
The relative size 
of the corresponding part in the cross section grows quadratically
in the energy $\sim E^2_{H}/m_H^2$. The size of the effect is governed by
the parameter $a_1$ which can be determined from the measurements 
of single Higgs-strahlung at JLC/NLC/TESLA. The  effect  is shown 
in Fig.~\ref{Edep} in which (a) the cross section difference and (b)
the cross section ratio for the double
Higgs production in $WW$-fusion are presented as a function of the
collider energy varied between 350 GeV and 5 TeV for a Higgs mass 
$m_H = 120$~GeV. The prediction of the Standard Model is compared
with the anomalous contributions for four values of the parameter 
$a_1 = \pm 0.05,\ \pm 0.1$.
We see from Fig.~\ref{Edep} that the dependence of the cross section 
on energy with the anomalous coupling $a_1$ is clearly stronger
than that of the SM. However, as observed from the cross section
ratio in Fig.~\ref{Edep}(b), the energy dependence is rather mild 
unless $a_1$ is as large as $a_1\sim 0.1$, and becomes more appreciable
at the CLIC energies $\sqrt s\sim 3$ TeV.
This is primarily due to the dominant contributions where
both Higgs bosons are radiated off $W$'s, that consequently
dilute the effect of the energy dependent
contribution from the Higgs derivative operators.

\section{Conclusions}
If the Standard Model is embedded 
as a low energy effective approach in a more comprehensive theory,
it is natural to include dimension-six operators 
as a first step in a systematic operator expansion for 
parameterizing the physics beyond the Standard Model.
We have discussed in this report the physics consequences of 
genuine dim-6 operators in the Higgs sector. 
The operators [with coefficients $a_1,\ a_2$ of order $v^2/\Lambda^2$] are not
subject to stringent constraints from electroweak precision data.
These operators renormalize the Higgs kinetic propagation 
and the Higgs mass parameter. Depending on the range of validity for the
expansion, they may even affect the vacuum stability
of the theory. Moreover, they can modify the couplings of the Higgs
boson to gauge bosons and the Higgs self-interactions.

We have studied the sensitivity to which those couplings can be probed at 
future $\ee$ linear colliders with $\sqrt s=500,\  800$ GeV, 
and in a subsequent second phase, with 3 and 5 TeV. 
We have found good accuracy in probing
$a_1$ and $a_2$, as presented in Figs.~\ref{senh_lc} and \ref{senhh_lc}.  
Some typical values of the
accuracy achievable at various collider energies 
are listed in Table~\ref{accu}, both in terms of $\Delta a_i$ and
$\Lambda$,
presenting the bounds to which dim-6 induced deviations from the
SM ($a_i=0$) can be probed.
The coefficient $a_1$ can be probed, and measured for non-zero
values, typically to better than the $0.01$ level,
and $a_2$ to the $0.1$ level.
We recall that $a_{i}\approx 0.05,\, 0.002$ for $f_{i} = 1$
leads to $\Lambda\approx 1,\, 5.5$ TeV. 
Thus the sensitivity reachable at future
linear colliders may be sufficient to explore the new effects as
parameterized by the dim-6 operators, at energy scales throughout
the threshold region of new strong interactions $\leq 4\pi v \sim 3$ TeV.

As a final remark in regard of a comparison with hadron colliders:
The LC precision tests for Higgs self-couplings are significantly
superior to the LHC tests. There is little chance to observe at the LHC 
the Higgs-pair events for a light Higgs boson with $m_H < 140$ GeV, 
which dominantly decays to $b\bar b$.  Only for a heavier Higgs boson
in the decay mode $H\to WW^*$ with leptons in the final state,
could the LHC, and thereafter improved by the VLHC, 
allow a first glimpse \cite{dtu} of the triple Higgs coupling.
\begin{table}[tbh]
\begin{tabular}{ l || l ||  c|c}
& $\sqrt s$ & $\Delta a_1\ (\Lambda\ \tev)$ & $\Delta a_2\ (\Lambda\ \tev)$\\
  \hline   \hline 
        & 500 GeV & ~0.0047~(3.6) & ~0.13~~(0.68)   \\
        & 800 GeV & ~0.0034~(4.2) & ~ 0.075~(0.89)   \\
  \raisebox{1.5ex}[0pt]{$m_H=120$ GeV}
        & 3 TeV   & ~0.0020~(5.5) & ~ 0.033~(1.4)   \\
        & 5 TeV   & ~0.0017~(6.0) & ~ 0.029~(1.4)  \\
\end{tabular}
\caption{Typical sensitivity $\Delta a_i$ to be reached as deviations from
the SM and the corresponding scales $\Lambda$ for $f_i=1$
at future linear colliders with 1 ab$^{-1}$. }
\label{accu}
\end{table}
%

\acknowledgments
We would like to thank Sasha Belyaev and Edward Boos 
for discussions, and P.L.~would like to thank A. Kusenko 
for a helpful correspondence.
This work was supported in part by the DOE under grants 
DE-FG02-95ER40896 and DOE-EY-76-02-3071, and in part by the 
Wisconsin Alumni Research Foundation.
T.H. and B.M. would also like to thank the DESY and LBNL 
Theory Groups, respectively, for their
hospitality while completing this paper.

\appendix
\section{The SM Higgs sector with dimension-six operators}\label{app_a}

In this appendix, we present some details of the effects of the genuine
dimension-six Higgs operators in extensions of the Standard Model.
These operators are non-renormalizable and can be induced by integrating
out heavy massive degrees of freedom in a more complete theory, as
commented in the text.

\subsection{The Higgs sector with genuine dimension-six operators}
The SM Lagrangian of the Higgs sector is of the form
\begin{eqnarray}
    {\mathcal{L}}_{SM}=|D_{\mu}\Phi|^2 -V_{SM},\quad
   V_{SM}=\mu^2|\Phi|^2 + \lambda|\Phi|^4 ,
\end{eqnarray}
where $\Phi$ is the Higgs doublet under the $SU_L(2)$ gauge group.  To
generate a local minimum away from zero field strength, $\mu^2$ must be
chosen negative, while $\lambda$ must be positive to guarantee the
stability of the system.  For simplicity we work in the unitary gauge.
After shifting the neutral scalar field with respect to its vacuum
expectation value to the physical Higgs field $H$, we write
\begin{equation}
    \Phi = {0 \choose (H+v)/\sqrt 2},
\end{equation}
where $v=\sqrt{-\mu^2/\lambda}$ is determined by minimizing $V_{SM}$.

At the dimension-six level, there are only two independent operators
that can be constructed purely from the Higgs field and that are not
constrained by existing data, as given in Eq.~(\ref{dim6text}). These
two operators generate kinetic and mass terms as well as three- and
four-point interactions:
\begin{eqnarray}
    \label{O1}
    {\mathcal{O}}_1 &=& {1\over 2}
        \partial_\mu H \partial^\mu H (v^2+2v H + H^2),\\
    {\mathcal{O}}_2 &=& -{1\over 24} (15 v^4 H^2 + 20 v^3 H^3
        + 15v^2 H^4).
    \label{O2}
\end{eqnarray}
The effective Higgs potential is the sum of $V_{SM}$ plus the
${\mathcal{O}}_2$ term,
\begin{equation}
    V_{eff}= \mu^2|\Phi|^2+\lambda|\Phi|^4+{f_2\over 3\Lambda^2}|\Phi|^6.
    \label{veff}
\end{equation}
This specific form may be interpreted as an expansion in $\Phi$,
truncated after the first anomalous term beyond the basic contributions
in the Standard Model.

\subsection{Electroweak symmetry breaking}
The electroweak symmetry breaking is induced by the non-zero vacuum
expectation value (vev) of the scalar field.  For curiosity we will
ignore the limited range of the truncated series and explore the
sexto-linear term in full generality. Conclusions restricted to the
``small field'' expansion have been elaborated in the main part of the
text.

Minimizing the potential Eq.~(\ref{veff}) with respect to $|\Phi|^2$
leads to the vev
\begin{equation}
    \langle |\Phi|^2\rangle \equiv {v^2\over 2}=
    \frac{-\lambda\pm \sqrt{\lambda^2-f_2\mu^2/\Lambda^2}}{f_2/\Lambda^2}
    \stackrel{\Lambda \gg \mu}{\longrightarrow}
        (\pm| \lambda| - \lambda)\frac{\Lambda^2}{f_2}
        \mp \frac{\mu^2}{2|\lambda|}
        \mp \frac{f_2 \mu^4}{8|\lambda|^3\Lambda^2}
        \pm \mathcal{O}\left(\frac{1}{\Lambda^4}\right).
    \label{vev}
\end{equation}
To obtain a real solution, a necessary condition is
$\lambda^2 \Lambda^2 \ge f_2\mu^2$.

\begin{figure}[tb]
    \includegraphics[scale=0.5]{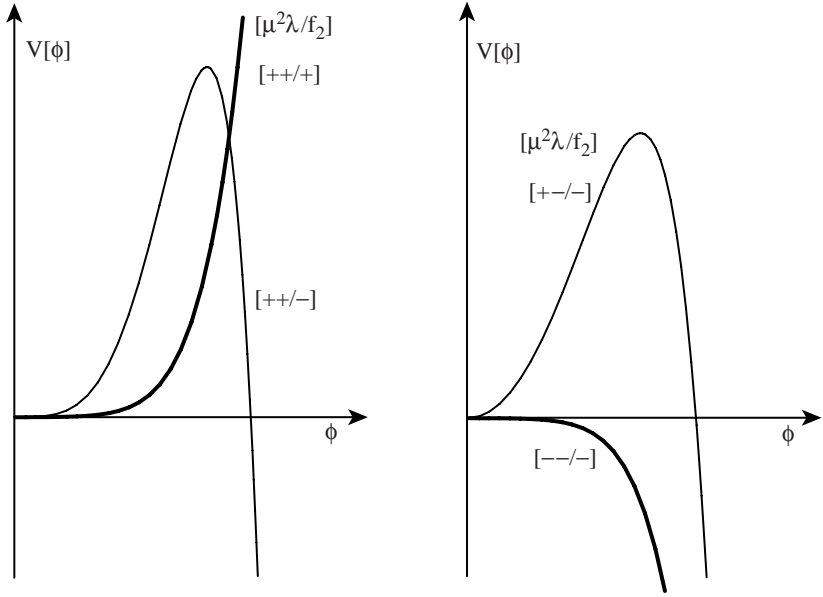}
    \caption{
        Higgs potential for the four no-electroweak symmetry breaking
        cases. 
The signs of the parameters $[\mu^2,\lambda,f_2]$ are given explicitly.
    }
    \label{badEWSB}
\end{figure}

\begin{figure}[tb]
    \includegraphics[scale=0.5]{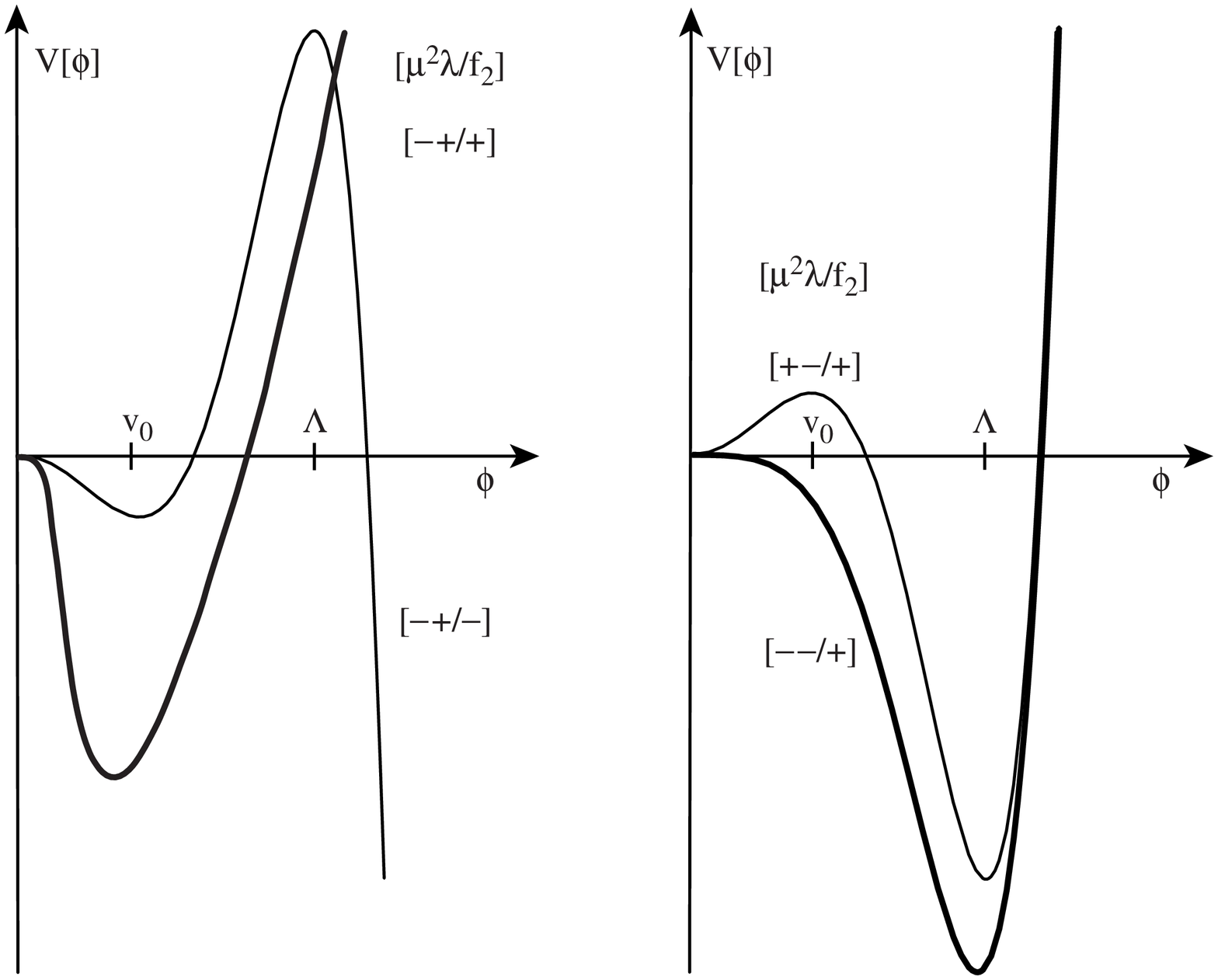}
    \caption{
        Higgs potential for the four cases with symmetry breaking
minima.
The signs of the parameters $[\mu^2,\lambda,f_2]$ are given explicitly.
The two cases in the left diagram connect smoothly to the Standard Model 
for $f_2\to 0$.
    }
    \label{goodEWSB}
\end{figure}

There are eight combinations of the sign choices for the three
parameters $\mu^2,\ \lambda$ and $f_2$. Four of them,
\begin{eqnarray}
    &&\mu^2>0,\quad \lambda>0,\quad f_2>0:
    \quad {\rm symmetric\ minimum\ at}\ \Phi=0;\nonumber\\ 
    &&\mu^2>0,\quad \lambda>0,\quad f_2<0:
    \quad {\rm symmetric\ minimum\ at}\ \Phi=0 \quad {\rm 
                                        and\ global\ maximum}\;\nonumber\\ 
    &&\mu^2>0,\quad \lambda<0,\quad f_2<0:
    \quad {\rm symmetric\ minimum\ at}\ \Phi=0 \quad {\rm  
                                        and\ global\ maximum}\;\nonumber\\ 
    &&\mu^2<0,\quad \lambda<0,\quad f_2<0:
    \quad {\rm no\ minimum}\nonumber
\end{eqnarray}
would not give a correct pattern for electroweak symmetry breaking, 
cf.~Fig.~\ref{badEWSB}.
We therefore discuss the other four configurations which can give rise
to spontaneous symmetry breaking indeed, cf.~Fig.~\ref{goodEWSB}.

\vskip 0.2cm
\noindent
\underline{$1.\ \mu^2<0,\quad \lambda>0,\quad f_2>0:$\ } 
The solution as given in Eq.~(\ref{vev}) is 
\begin{equation}
    {v^2\over 2}= \frac{\lambda\Lambda^2}{f_2}
    \left(\sqrt{1+{f_2|\mu^2|\over \lambda^2\Lambda^2}}-1 \right).
    \label{caseii}
\end{equation}
It is easy to check that $V_{eff}(v^2/2)<0$ and that this is a global
minimum. Only far away from the Standard Model 
[realized for $\Lambda \gg |\mu|$], 
the system in the ground state would be determined by the
parameter $\Lambda$ with $v^2/2 \sim \sqrt{f_2} \Lambda |\mu|$.
However, if the sexto-linear term acts, for large $\Lambda$, as a
correction to the SM terms, the system is primarily governed by $\mu^2$
and $\lambda$, as in the Standard Model.  For the large-$\Lambda$
expansion:
\begin{equation}
    v^2 = v_0^2 - \frac{f_2 v_0^2}{4\lambda\Lambda^2}
    \label{vev1}
\end{equation}
with ${v_0}^2 = |\mu^2|/\lambda$.   

\vskip 0.2cm
\noindent
\underline{$2.\ \mu^2<0,\quad \lambda>0,\quad f_2<0:$\ }
In this case the Higgs potential generates a minimum away from
zero and breaking the symmetry, but in addition a maximum, for
large $\Lambda$ far away from the minimum. The expansion in $\Lambda$,  
\begin{equation}
    {v^2\over 2}\approx \frac{\lambda\Lambda^2}{|f_2|}
    \left[\pm\left(1-{|f_2\mu^2|\over 2\lambda^2\Lambda^2}-{1\over 8}
    \left({f_2\mu^2\over \lambda^2\Lambda^2}\right)^2\right)+1 \right],
\end{equation}
the two solutions for the {\it loci} of the extrema are given by the
values: 
\begin{eqnarray}
    {v^2_1}\approx {v^2_0}\left(1+\frac{|f_2|v^2_0}{4\lambda\Lambda^2} 
        \right), \quad 
    {v^2_2}\approx {v^2_0}\left(\frac{4\lambda\Lambda^2}{|f_2|v^2_0}-1\right)
        \approx \frac{4\lambda\Lambda^2}{|f_2|}.
\end{eqnarray}
Clearly, the first solution $v_1$ is of the SM form with an order
$v^2_0/\Lambda^2$ correction, and it is a deeper minimum. 

The second solution $v_2$, however, is a maximum. 
If the coefficient $f_2$ is negative, then the potential is not bounded
from below and it leads to an ultimately unstable configuration at large
$|\Phi|$ \cite{jose}. 
In fact, if $V_{eff}$ in Eq.~(\ref{veff}) holds for very large
$\Phi$ [{\it i.e.}, $\Phi \sim O(100 \Lambda)$] then the tunneling rate
to the unbounded vacuum is extremely large.  However, for such large
$\Phi$ the effective theory will no longer be valid and
higher-dimensional terms such as $\Phi^8/\Lambda^4$ in the $1/\Lambda^2$
series will become important, so that the stability of the system can be
restored again.

The instability of the system 
can easily be proved. If Eq.~(\ref{veff}) holds up to very large
$\Phi$, the tunneling rate is in the ``thick-wall'' regime, which can be
estimated by the methods described in~\cite{linde}.  We sketch the
calculation for the simplified case $\mu^2=0$, since the barrier height
and width are mainly controlled by $\lambda$ and $f_2/\Lambda^2$, and
the shift in $\mu$ is inessential.  In that case, the potential is
\begin{equation}
    V(H) = \frac{\lambda}{4} H^4 - \frac{\kappa^2}{6} H^6,
\end{equation}
where $\Phi = H/\sqrt{2}$ and $\kappa^2 = -f_2/4\Lambda^2$.  The
tunneling is described by the minimum action $O(4)$ symmetric solution
characterized by $r=\sqrt{\vec{x}^2 + t^2}$ for the equation of motion
for $H$ in Euclidean space.  It is convenient to introduce dimensionless
variables $\psi$ and $R$ by
\begin{equation}
    H = \sqrt{\frac{3 \lambda}{2}}\ \frac{\psi}{\kappa}\ , \ \ \  r =
    \sqrt{\frac{2}{3}}\ \frac{\kappa R}{\lambda}.
\end{equation}
Then the tunneling rate per unit volume is proportional to $\exp{(-S)}$,
with the minimum action configuration $S(\phi)$ given by $S(\phi) =
S(\psi)/\lambda$, where
\begin{equation}
    S(\psi) = 2 \pi^2 \int_0^\infty R^3 dR \left[\frac{1}{2}
     \left(\frac{d \psi}{dR}\right)^2
       + \frac{1}{4} \psi^4 (1-\psi^2)\right]
\end{equation}
and $\psi$ is the solution to
\begin{equation}
    \frac{d^2 \psi}{d R^2} + \frac{3}{R} \frac{d \psi}{d R}
        = \psi^3 (1-\frac{3}{2}\psi^2),
\end{equation}
subject to the boundary conditions $\psi'(0) = 0$, and $\psi(\infty) =
0$.  $S(\phi)$ is independent of $\kappa$, due to a compensation between
the energy density in the bubble and its size.  A numerical solution
yields the extremely small action $S \sim 0.00084/\lambda$, which is
much smaller than the value $S \agt 400$~\cite{kusenko} needed to ensure
less than one transition during the age of the observable universe. The
essentially instantaneous decay is due to the extremely steep fall off
of the potential for large $\psi$.  However, as noted above, the
solution is only valid if Eq.~(\ref{veff}) is valid up to $\psi \agt
\psi(0)\sim 137$, corresponding to $\phi/\Lambda \sim
335\sqrt{\lambda/|f_2|}$.  The negative sign $f_2 < 0$ can therefore
only be allowed, if the potential Eq.~(\ref{veff}) is replaced by a more
complete expression before $H$ reaches that large a value.

\vskip 0.2cm
\noindent
\underline{$3.\ \mu^2>0,\quad \lambda<0,\quad f_2>0:$\ } 
In this case, there three extrema that can be readily read off 
from Eq.~(\ref{vev}) for large $\Lambda$:
\begin{eqnarray}
    \label{sol1}
    {v^2_1} = 0, \quad
    {v^2_2}\approx {v^2_0}\left(1+\frac{f_2v^2_0}{4|\lambda|\Lambda^2} 
    \right), \quad 
    {v^2_3}\approx {v^2_0}\left(\frac{4|\lambda|\Lambda^2}{f_2v^2_0}-1\right)
    \approx \frac{4|\lambda|\Lambda^2}{f_2}.
\end{eqnarray}
It turns out that following the symmetric minimum at
$v_1 = 0$, there is a maximum at $v_2$, and there is a minimum
at $v_3$, either above or below $V=0$. This scenario is undesirable 
from a theoretical point of view in any case.
The reason is that the feasible vacuum $v_3$ is determined by 
$|\lambda|\Lambda^2/f_2$, and if we consider the dim-6 operators 
as corrections to the SM potential,
we wish the vacuum not to be determined by the correction term or
by the cutoff scale $\Lambda$, but rather to recover
the SM by taking the decoupling limit $\Lambda\to \infty$, which
would not be achieved if we accept this vacuum.

\vskip 0.2cm
\noindent
\underline{$4.\ \mu^2<0,\quad \lambda<0,\quad f_2>0:$\ }
In this case, following the symmetric maximum at $|\Phi|=0$, 
the vacuum solution is realized for large $\Lambda$:
\begin{equation}
    {v^2}\approx
    {v^2_0}\left(\frac{4|\lambda|\Lambda^2}{f_2v^2_0}+1\right)
    \approx \frac{4|\lambda|\Lambda^2}{f_2}.
    \label{caseiv}
\end{equation}
Again, as discussed above, the solution is determined by the cutoff 
scale $\Lambda$, and would not be desirable to pursue.

\subsection{Corrections to Higgs boson couplings}
We first note that the dim-6 operator ${\cal O}_1$ corrects the kinetic
terms of the Higgs boson propagation, as seen from Eq.~(\ref{O1}), and
the full kinetic terms become
\begin{equation}
    \mathcal{L}_{kin} = {1\over 2}\partial_\mu \phi \partial^\mu \phi +
    {1\over 2}{f_2 v^2\over \Lambda^2}\partial_\mu \phi \partial^\mu \phi.
    \label{kin}
\end{equation}
We thus re-scale the field $\phi$ to define canonically normalized
Higgs field $H$
\begin{equation}
    \phi = N H,\quad {\rm with}\quad 
    N={\left(1+ a_1 \right)}^{-\frac{1}{2}},
    \label{N}
\end{equation}
with $a_1 = f_1 v^2 /{\Lambda^2}$ used hereafter.

With this field-redefinition, we find from the effective potential of
Eq.~(\ref{veff}) the physical Higgs boson mass 
\begin{eqnarray}
    \label{mh}
    m_H^2 &=& N^2\left(\mu^2 + 3\lambda v^2 + 
    \frac{5f_2v^4}{4\Lambda^2}\right)\\
          &\approx& 2\lambda v^2\left(1 - 
               \frac{f_1v^2}{\Lambda^2} +
               \frac{f_2v^2}{2\lambda\Lambda^2}\right), \nonumber
\end{eqnarray}

As for the gauge-boson-Higgs couplings, each Higgs field receives
a correction factor $N \approx 1 - a_1/2$, leading to the interactions
with the linearized anomalous coupling
\begin{eqnarray}
    \label{VVH}
    \mathcal{L}_M &=& (M_W^2W^+_\mu W^{-\mu}+{1\over 2}M_Z^2Z_\mu Z^\mu)
    (N{2 H\over v}+N^2{H^2\over v^2})\\
    &\approx& (M_W^2W^+_\mu W^{-\mu}+{1\over 2}M_Z^2Z_\mu Z^\mu)
    \left((1-{a_1\over 2}){2H\over v}+(1-a_1){H^2\over v^2}\right).
\end{eqnarray}

The Higgs self-interactions are from $V_{SM}$, (\ref{O1}) and (\ref{O2})
and we write them in terms of the three-point and four-point couplings
\begin{eqnarray}
    \nonumber
    \mathcal{L}_{H^3} + \mathcal{L}_{H^4} &=& -\lambda v N^3( H^3-{a_1H\over 
                                                                      \lambda}
    {\partial_\mu H \partial^\mu H\over v^2} + 
    {5a_2\over 6\lambda} H^3) \\ 
    \nonumber
    && -{\lambda\over 4} N^4(H^4 - {2a_1 H^2\over \lambda}
    {\partial_\mu H \partial^\mu H\over v^2} + 
    {5a_2\over 2\lambda} H^4)\\
    \nonumber
    &\approx& -{m_H^2\over 2 v}
    \left( (1-{a_1\over 2}+{2a_2\over 3}{v^2\over m_H^2}) 
    H^3-{2a_1 H \partial_\mu H \partial^\mu H\over m_H^2} \right) \\ 
    \label{Hs4}
    && -{m_H^2\over 8v^2} \left( (1-a_1 + {4a_2v^2\over m_H^2}) H^4 -
    {4a_1H^2\partial_\mu H \partial^\mu H\over m_H^2} \right).
\end{eqnarray}
The Feynman rules for the Higgs self-coupling vertices are given as
\begin{eqnarray}
    \label{3hfeynman}
     HHH:&&
    -i{3m_H^2\over v}
    \left( (1-{a_1\over 2}+{2a_2\over 3}{v^2\over m_H^2}) +
    {2a_1  \over 3m_H^2} \sum^3_{j<k}\ p_j \cdot p_k\right) \\ 
     HHHH:&&  
     -i{3m_H^2\over v^2} \left( (1-a_1 + {4a_2v^2\over m_H^2}) +
    {2a_1 \over 3m_H^2}\sum^4_{j<k}\ p_j \cdot p_k\right).
\end{eqnarray}


\end{document}